\documentclass{article}

\usepackage{arxiv}

\usepackage[utf8]{inputenc} 
\usepackage[T1]{fontenc}    
\usepackage{hyperref}       
\usepackage{url}            
\usepackage{booktabs}       
\usepackage{amsfonts}       
\usepackage{nicefrac}       
\usepackage{microtype}      
\usepackage{lipsum}		
\usepackage{graphicx}
\usepackage{natbib}
\usepackage{doi}

\usepackage{subcaption}
\usepackage{amsmath}

\title{Unraveling the Geography of Infection Spread: Harnessing Super-Agents for Predictive Modeling}

\author{ \href{https://orcid.org/0000-0001-6105-5089}{\includegraphics[scale=0.06]{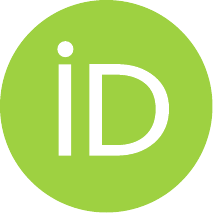}\hspace{1mm}Amir Mohammad Esmaieeli Sikaroudi}\\
	Computer Science Department\\
	University of Arizona\\
	Tucson, AZ 85721 \\
	\texttt{amesmaieeli@arizona.edu} \\
	\And
	\href{https://orcid.org/0000-0001-9834-4132}{\includegraphics[scale=0.06]{orcid.pdf}\hspace{1mm}Alon Efrat} \\
	Computer Science Department\\
	University of Arizona\\
	Tucson, AZ 85721 \\
	\texttt{alon@arizona.edu} \\
    \And
	\href{https://orcid.org/0000-0002-6758-515X}{\includegraphics[scale=0.06]{orcid.pdf}\hspace{1mm}Michael Chertkov} \\
	Program in Applied Mathematics \& Department of Mathematics\\
	University of Arizona\\
	Tucson, AZ 85721 \\
	\texttt{chertkov@arizona.edu} \\
}

\renewcommand{\shorttitle}{\textit{arXiv} Template}

\hypersetup{
pdftitle={A template for the arxiv style},
pdfsubject={q-bio.NC, q-bio.QM},
pdfauthor={David S.~Hippocampus, Elias D.~Striatum},
pdfkeywords={First keyword, Second keyword, More},
}

\begin{document}
\maketitle

\begin{abstract}
    Our study presents an intermediate-level modeling approach that bridges the gap between complex Agent-Based Models (ABMs) and traditional compartmental models for infectious diseases. We introduce "super-agents" to simulate infection spread in cities, reducing computational complexity while retaining individual-level interactions. This approach leverages real-world mobility data and strategic geospatial tessellations for efficiency. Voronoi Diagram tessellations, based on specific street network locations, outperform standard Census Block Group tessellations, and a hybrid approach balances accuracy and efficiency. Benchmarking against existing ABMs highlights key optimizations. This research improves disease modeling in urban areas, aiding public health strategies in scenarios requiring geographic specificity and high computational efficiency.
 
\end{abstract}


\section{Introduction}
In the realm of epidemiology, conventional models such as the Susceptible-Infectious-Recovered (SIR) compartmental framework, originating from the papers of Ross \cite{1910Ross}, Kermack \cite{1927Kermack}, Anderson \cite{1991Anderson} and many others, offer a highly aggregated view of disease spread within populations. While these models are valuable for certain applications, they often simplify interactions and behaviors to the point where individual--level details are lost. In contrast, our approach serves as an intermediary, striking a balance between the complexity of Agent-Based Models (ABMs) \cite{perez_agent-based_2009, auchincloss_new_2008, epstein_coupled_2008,epstein_modelling_2009,nianogo_agent-based_2015} and the extreme aggregation of compartmental models \cite{1910Ross,1927Kermack,1991Anderson}.

We introduce 'super-agents' to enhance the granularity of our simulations, enabling us to bridge the gap between ABMs and compartmental models. This 'super-agent' methodology significantly preserves individual-level interactions and behaviors, allowing us to explore disease spread within urban areas with a high degree of fidelity while still benefiting from computational efficiency. By embracing the strengths of both ABMs and compartmental models, our research offers a novel perspective on infectious disease modeling that can cater to a broader range of scenarios and research questions.

Now, returning to the core concept of agent-based simulation, it involves the creation of virtual entities known as 'agents,' each equipped with individual characteristics, behaviors, and decision-making abilities. These agents interact with each other and their environment, giving rise to emergent behavior at the system level. Agent-based simulation finds applications in numerous fields, including sociology, economics, biology, ecology, computer science, and urban planning, among others. It serves as a virtual laboratory for exploring complex scenarios and conducting 'what-if' analyses, making it an indispensable tool for decision-making, policy formulation, and understanding the dynamics of complex systems.

An agent can represent a wide range of entities, such as individuals, animals, organizations, or even abstract entities like ideas or market forces. The primary advantage of agent-based simulation lies in its ability to capture the intricacies of interactions and behaviors within a system, making it particularly suitable for studying systems with non-linear, dynamic, and uncertain characteristics.

To facilitate the simulation process, agents are programmed with predefined rules or algorithms that dictate their actions and responses to various stimuli and events. As the simulation progresses, agents continuously adapt their behaviors based on internal states, external inputs, and interactions with other agents. Consequently, the aggregate behavior of the system emerges from the cumulative actions of individual agents, allowing researchers to gain insights into the collective behavior and outcomes of the modeled system.

This sets the stage for the classic Agent-Based Modeling (ABM), which forms the core focus of our manuscript. Our exploration centers on the integration of ABM and geography -- a powerful and innovative approach to understanding the complex dynamics of real-world systems within a spatial context. ABM is a computationally heavy approach to study pandemic and it requires integration of various datasets and dynamics. This study discusses methods to facilitate large-scale simulations while minimizing the quality decline. We investigate geographical abstraction and reduction of the number of agents. However, to make such abstraction possible and rigorous, we need to introduce an ABM model that extends the models from the literature to present detailed geography, mobility, and disease transmission in large-scale.

In our endeavor to advance infectious disease modeling, we recognize the distinct strengths and limitations inherent to both ABMs and traditional compartmental models. ABMs excel in capturing the detailed behaviors and interactions of individuals within a population, offering a high-resolution lens through which complex systems can be examined. This granularity enables the exploration of scenario-specific dynamics and the effects of individual-level interventions, which are often overlooked in more aggregated modeling approaches. Conversely, compartmental models provide a streamlined, macroscopic view of disease dynamics, allowing for the efficient analysis of population-wide trends and outcomes. Such models are valued for their mathematical simplicity and computational efficiency, facilitating the examination of long-term epidemic trends and the impact of broad-scale public health measures.

The Agent-in-Cell (AIC) model introduced in this manuscript seeks to harness these complementary strengths by integrating the detailed, micro-level insights afforded by ABMs with the broader, systemic perspective characteristic of compartmental models. This hybrid approach enables our research to cater to a wider range of scenarios and research questions, from the intricate mapping of local outbreaks to the overarching patterns of pandemic spread. By doing so, we offer a novel perspective on infectious disease modeling that bridges the gap between the nuanced, agent-specific behaviors captured by ABMs and the generalized, population-level insights derived from compartmental models. This integrative model aims not only to enrich our understanding of infectious disease dynamics but also to enhance the robustness and applicability of modeling efforts across diverse epidemiological contexts.

\section{Background and Literature}

\label{sec:BackgroundLiterature}

While the integration of the ABM with geographic information systems (GIS) has been explored by researchers such as A. Crooks, our approach seeks to extend this foundational work by incorporating open-source data and mathematically justified tesselation approach to resolving geographical details. Crooks' significant contributions have laid the pre-COVID groundwork for the fruitful intersection of ABM and geography, providing essential insights and methodologies that have informed our research. Specifically, we draw upon Crooks' methods of utilizing GIS data within ABMs to enhance the spatial accuracy and relevance of our simulations, as outlined in his comprehensive overview on GIS and agent-based modeling and his detailed examination of epidemiological models before the COVID-19 pandemic \cite{crooks2014agent}. Our work aims to build upon these efforts.

Beyond this earlier work on expressing geography via ABM, the realm of ABM abstraction and simulation optimization remains relatively unexplored. Rhodes et al. \cite{rhodes2016reducing} delved into the potential of modeling complex systems with a reduced number of agents, examining interaction rates, iteration durations, and messaging overhead. Similarly, Tregubov et al. \cite{tregubov2021optimization} investigated sub-sampling agents to represent groups and teams in a social media model, concluding that the representative agent approach effectively reduces runtime while maintaining reasonable performance levels.

Medical simulations have also proposed ABM abstraction with a reduced number of agents. Shirazi et al. \cite{shirazi2014adaptive} introduced super-agents to simulate the blood coagulation process with fewer computational resources. These super-agents adhere to aggregated stochastic rules, as opposed to individual rules for each agent. In another study by Sarraf et al. \cite{sarraf2013abstraction}, interaction abstraction was explored for blood coagulation, suggesting that the model could learn appropriate abstraction levels by assessing interaction validity, with reversions for invalid abstractions.


A pandemic is characterized by the widespread outbreak of a disease that exhibits specific features, including extensive geographical spread, low levels of immunity within the population, and a rapid rate of transmission \cite{morens2009pandemic}. Historically, pandemics are often linked to interactions between humans and animals capable of transmitting pathogens to humans. Morens et al. conducted a historical analysis of pandemics, exploring various causative factors that can lead to such global health crises \cite{morens2020pandemic}. Agent-based models (ABMs) serve as a pivotal tool in examining the dynamics of pandemics, facilitating the exploration of potential mitigation strategies. For example, FlueTE, an ABM specifically designed for influenza, simulates the pandemic's dynamics to evaluate the effectiveness of pharmaceutical interventions and social distancing measures \cite{chao2010flute}. The recent COVID-19 pandemic has further underscored the value of ABMs, enabling the assessment of various intervention policies aimed at minimizing the impact on public health. Extensive research has focused on analyzing the impact of COVID-19 lockdowns on human behavior, primarily within urban contexts. The pandemic has brought about varied shifts in people's daily routines, notably affecting aspects such as public transportation preferences \cite{aloi2020effects} and food shopping habits \cite{philippe2021child}. Researchers have leveraged the SafeGraph mobility dataset \cite{SafeGraphDataConsortium} to probe into diverse dynamics including economic and social impacts due to the pandemic. This dataset records the daily movement of individuals from specific census tracks (residential areas) to commercial locations. For instance, Goolsbee and Syverson \cite{goolsbee2021fear} examined the business impact by analyzing pre-pandemic and pandemic visit rates, finding that while legal restrictions caused only a minor 7\% reduction in customer traffic, most of the decline was attributed to individual COVID concerns \cite{goolsbee2021fear}.

Furthermore, SafeGraph data has enabled investigations into the effects of socio-economic factors on mobility patterns during the pandemic. Weill et al. \cite{weill2020social} demonstrated that mobility reduced in high-income areas while rising in low-income neighborhoods \cite{weill2020social}. In a related vein, Yan et al. \cite{yan2021risk} explored the interplay between face mask mandates and social behaviors using SafeGraph data, studying the connection between stay-at-home durations and visits to commercial establishments \cite{yan2021risk}. Additionally, there have been attempts to model COVID-19 dynamics using differential equations with parameters inferred from SafeGraph data \cite{chen2020state,chang2021mobility}.

A pivotal aspect of this paper involves modeling mobility through a bipartite graph structure, wherein the nodes are categorized into two types: residential areas, referred to as Census Block Groups (CBGs), and commercial locations termed \textit{Points of Interest} (POIs). Utilizing the SafeGraph dataset, Chang and colleagues \cite{chang2021mobility} successfully estimated the hourly travel counts of individuals or agents moving from CBG $i$ to POI $j$.

Since the onset of the COVID-19 pandemic, a plethora of new open-source ABM tools have emerged. Among these, Covasim stands out, offering pandemic transmission modeling, intervention scenarios, and a versatile modeling library \cite{kerr2021covasim}. Shamil and colleagues \cite{shamil2021agent} have developed an agent-based simulation incorporating non-pharmaceutical interventions like lockdowns to control the reproduction number of COVID-19. However, their approach doesn't account for geographic heterogeneity, although they do differentiate agents' activities based on their occupations.

Similarly, Silva and team \cite{silva2020covid} considered various lockdown scenarios and assessed their economic implications but omitted geographical information. DESSABNeT is another notable ABM designed to simulate individual behavior and infection dynamics, though it lacks the geospatial trajectories of agents' activities \cite{stapelberg2021discrete}. Gupta and colleagues \cite{gupta2020covi} delved into the impact of COVID-19 contact tracing on virus spread and analyzed economic consequences of interventions in Montreal, relying on aggregated geographical data.

OpenABM-COVID-19, as outlined by Hinch et al. \cite{hinch2021openabm}, has concentrated on medical aspects like different strains and vaccination patterns. Although geospatial information isn't a focal point, it emphasizes demographic data pertinent to geo-aggregated medical modeling. A distinctive approach was presented by Mahmood et al. \cite{mahmood2020facs}, who implemented FACS, an ABM accounting for the geospatial inhomogeneity of agent mobility in London. They evaluated their model by comparing the number of hospitalization between historical data and their model's output. FACS-CHARM builds upon the FACS simulator by integrating hospital simulations, including aspects like ward management and intensive care unit operations. This extension offers a more comprehensive view on the utilization of medical services and hospital bed occupancy \cite{anagnostou2022facs}.

In response to valuable insights from recent systematic reviews, notably the comprehensive analysis of ABMs in COVID-19 research highlighted by  Lorig et al. \cite{lorig2021agent}, we would like to acknowledge the diverse and heterogeneous application of ABMs and GIS data in pandemic modeling. Recognizing the identified gaps in data representation and the scarcity of open-source implementations, our approach uniquely integrates detailed geographic tessellation and provides an open-source framework. This methodology not only enhances the granularity and accuracy of pandemic simulations but also promotes transparency and reproducibility in the field.

\subsection*{Context within Broader Modeling Efforts}

In this subsection, we provide a subjective and brief review of various modeling approaches, both from our (extended) team and others, related to pandemic modeling, with a particular focus on the COVID-19 pandemic. Given the global impact of COVID-19, which has affected every country to varying degrees, it has become the most extensively researched pandemic in history. Consequently, we have chosen to focus our methodology on COVID-19, leveraging the vast array of data available to enhance the robustness and applicability of our approach.

The classical compartmental models, such as the widely used SIR (Susceptible-Infectious-Removed) and age-of-infection models, have long been staples in epidemiology \cite{1910Ross,1927Kermack,1991Anderson,2000Hethcore}. While these models excel in post-factum explanations, they exhibit significant limitations when it comes to predicting ongoing and future disease outbreaks with the required accuracy. These limitations are particularly evident in recent COVID-19 inspired work addressing temporal variability \cite{tkachenko_stochastic_2021,tkachenko_time-dependent_2021}, geographical heterogeneity \cite{hinch2021openabm,gupta2020covi,mahmood2020facs,gomez2020infekta,chang_mobility_2020,Chen_2020}, and the impact of super-spreader events, where a small number of individuals can lead to widespread infection \cite{wang_inference_2020,2020Endo}.

The shortcomings of classical models can be attributed to several factors. Firstly, their compartmental nature, while groundbreaking at its inception, becomes a limitation when compartments are overly large. To model a city effectively, especially one operating under pandemic restrictions, we must introduce compartments that align with fine-grained geographical resolutions. Attempting to describe these smaller, geo-separated compartments using aggregated characteristics, such as the total number of susceptible, infectious, and removed individuals per compartment, remains incomplete if the model remains deterministic. Consequently, classical models tend to overlook spatio-temporal fluctuations, which have a significant impact on disease dynamics. Secondly, model calibration historically relied on extensive data training, a practice that has become feasible only in the past decade with the advent of Deep Learning in AI and Data Science. This has opened new avenues for epidemiology and other scientific disciplines to harness advanced technology.

\section*{Methodology}

\subsection*{Agent-Based model}

Quite a number of ABM implementations for pandemics exist, each offering varying levels of detail. Our ABM, tailored to assess the impact of geospatial tessellation, is dubbed the Agent-in-Cell (AIC) model. AIC uniquely assigns each agent to a tessellation cell. We've employed established ABM methodologies, especially those relevant to epidemiology, in crafting the AIC, as discussed in the background and literature section. Our AIC draws inspiration from the framework introduced by Shamil et al. \cite{shamil2021agent}. This section delves into the key facets of AIC, with additional specifics elaborated upon in the supplementary material.

\begin{figure}[h]
    \captionsetup{justification=centering}
    \centering
    \includegraphics[scale=0.31]{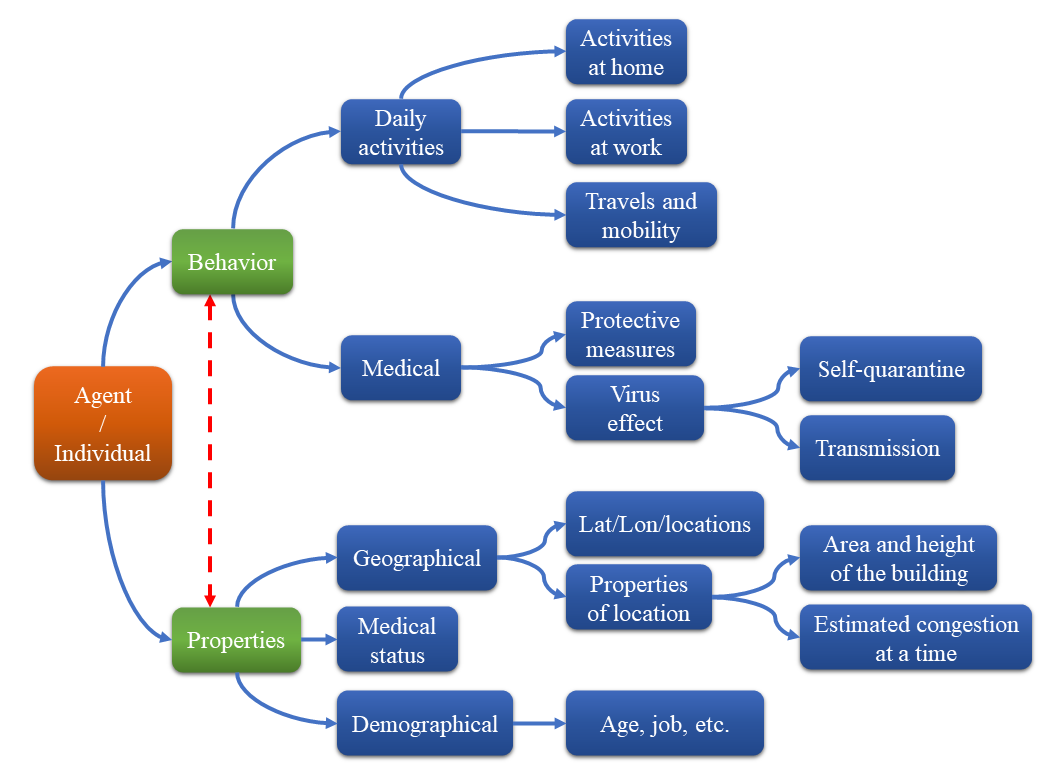}
    \caption{Diagram illustrating the fundamental components of our customized ABM model of Pandemic - the Agent-in-Cell (AIC) model.
    }
    \label{fig:ABM_sche}
\end{figure}

\begin{figure}[h]
    \centering
    \includegraphics[scale=0.36]{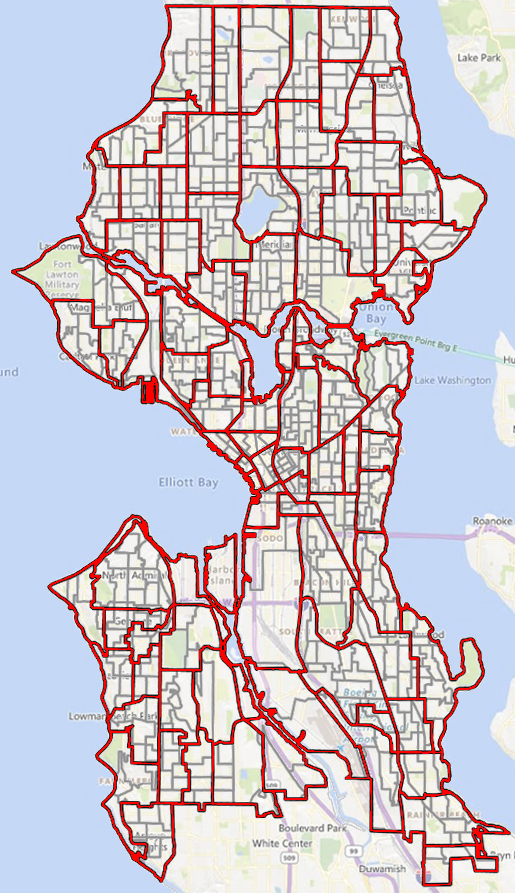}
    \caption{This map of Seattle depicts census tracts with red boundaries and census block groups with gray boundaries (Source: \cite{seattleCensusTractMap2010}).}
    \label{fig:seattleCBGs}
\end{figure}

Within AIC, each agent adheres to mobility regulations, environmental interaction protocols (both at home and work), and interpersonal interaction guidelines, potentially leading to transmission events during trips. Figure \ref{fig:ABM_sche} offers an overview of the interplay between behavioral patterns and related properties within the AIC framework. It's crucial to highlight that all these factors are linked to geographical locations. The spread of infection is treated as a stochastic process, occurring at each time step with a Poisson rate tailored to the agent's situation and interactions with peers. Agents can be either actively engaged at specific Points of Interest (POIs) or within distinct city areas, referred to as tessellation cells. Among the available tessellation options, the Census Block Group (CBG) is highlighted. An illustration of the CBG-based tessellation is presented in Figure \ref{fig:seattleCBGs} for the city of Seattle. Rooted in U.S. federal geographical identifiers, CBGs correspond to census areas, which inherently implies smaller CBG areas in densely populated regions and larger ones in sparser locales. Each CBG typically hosts a population ranging from 600 to 3,000 \cite{us2021glossary}.

\subsection{Daily trip Activities}

Apart from considering agents' activities at their residences and workplaces, we also incorporate their daily travel-related engagements, encompassing activities such as shopping and visits to various Points of Interest (POIs). These daily excursions bear significance as they introduce a geographical dimension to the model. An essential modeling hurdle lies in accurately representing the mobility patterns of each individual agent. Understanding their destinations and interactions is crucial for effectively tracking pandemic transmission. However, acquiring such detailed information poses privacy concerns. To address this challenge, we turn to the SafeGraph dataset \cite{SafeGraphDataConsortium}, offering a partial solution to our data needs.

The SafeGraph dataset from \cite{SafeGraphDataConsortium} offers a valuable resource for tracking mobility patterns while circumventing privacy concerns. It achieves this by furnishing (a) aggregated counts of visits to specific Points of Interest (POIs) and (b) aggregated Census Block Groups (CBGs) as the origins of these trips. This dataset provides insights into the movement from CBG $i$ to location $j$, with a time granularity of an hour, and also includes the classification of destination type, such as accommodation, food service, schools, and more. The data is collected by monitoring cellphone users, acknowledging that it might not capture the entire population and could potentially introduce biases. Despite these limitations, it remains a widely employed open dataset for related research purposes.

\subsection{Activities at home and at work}

We adhere to the standard ABM procedure by initiating the simulation with the creation of random household instances. This is accomplished based on the population density of the corresponding Census Block Groups (CBGs) and the household size statistics retrieved from the US Census Bureau \cite{uscensusbureau2021}. Each employed agent is then assigned a workplace location, selected randomly according to the relevant CBG-specific statistical input. Conversely, non-working agents are assumed to remain at home. An agent's work-related activities are grouped together, which facilitates the evaluation of new infection transmission between colleagues who share a common work location. This transmission occurs probabilistically and is influenced by the time spent at the same workplace. Job-related trips are discerned based on the information furnished by SafeGraph, particularly the daytime CBG. Notably, if an agent's daytime CBG matches their home CBG, it is inferred that the agent is working from home.

\subsection{Disease Transmission Modeling}

The efficacy of contact between two agents hinges on both their physical proximity and the chosen proximity metric. To account for uncertainty and stochastic effects, it is logical to introduce a proximity measure influenced by an agent's activity group. A fundamental contribution of this manuscript lies in the development of a fitting proximity measure for travel scenarios. Infection transmission occurs as agents interact with infected individuals or come into contact with contaminated surfaces. We address these mechanisms through parameterized models, extracting pertinent parameters from previous research. For instance, parameters such as the typical duration of pathogen viability on surfaces and the likelihood of infection through surface contact are adopted from Mwalili et al. \cite{mwalili2020seir}. Agent interactions are characterized with granular detail, including considerations of the geometric attributes of the agent's dwelling or visited locations. We draw inspiration from Agrawal and Bhardwaj \cite{agrawal2021probability} and Mittal et al. \cite{mittal2020mathematical}, leveraging their reported estimates on transmission probabilities. Furthermore, we incorporate insights from Agrawal and Bhardwaj \cite{agrawal2021probability} regarding various aspects of disease transmission, including aerosol dynamics, underlying fluid mechanics, and probability estimations based on the distance between infected and healthy individuals.

\begin{figure}[h]
    \captionsetup{justification=centering}
    \centering
    \includegraphics[scale=0.36]{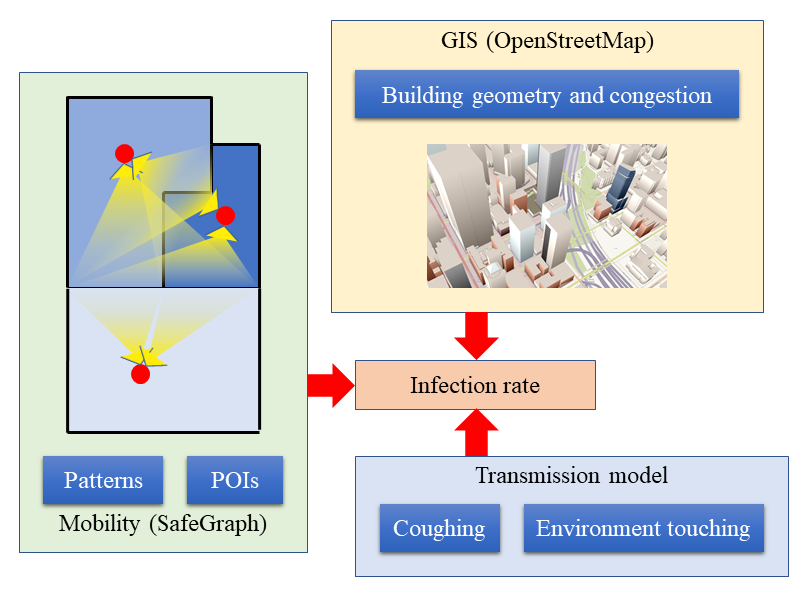}
    \caption{
     Schematic view of the infection model.}
    \label{fig:ABM_infectionModel}
\end{figure}

\begin{figure}[h]
    \captionsetup{justification=centering}
    \centering
    \includegraphics[scale=0.36]{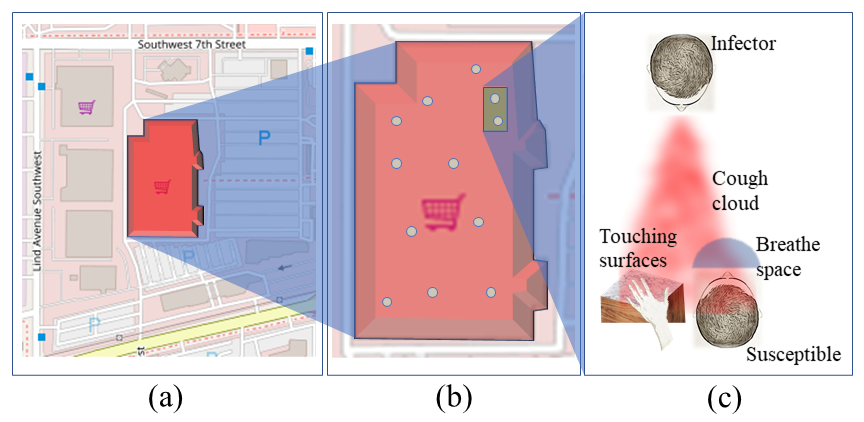}
    \caption{The infection model for trips. (a) Building location and geometry. (b) Estimated congestion. (c) Physical transmission models (base on \cite{agrawal2021probability})}
    \label{fig:ABM_infectionBuilding}
\end{figure}

The various inputs integrated into our infection model are illustrated in Figure \ref{fig:ABM_infectionModel}, while more specific information regarding the physical modeling of infection spread within a POI is provided in Figure \ref{fig:ABM_infectionBuilding}.
 
We assume a uniform distribution of people within a building and thus introduce the density of people within building (POI) $i$ at time $t$ as $\delta(t,i)$. This density is influenced by $N_{t_i}$, the estimated number of people in a POI at time $t$, and by the building geometry derived from the OpenStreetMap (OSM) dataset \cite{OpenStreetMapCitation}. While many POIs like groceries, schools, and offices are likely present on OSM, for unidentified POIs, we estimate them based on the average size of buildings in the area. The building geometry is assumed to be dependent on the land area of the building, $A_i$, and the number of floors, $F_i$. $N_{t_i}$ is extracted from the SafeGraph database, and then the expected distance between a pair of agents at time $t$, co-located in building/POI $i$, is estimated as follows:
\begin{align}
 \delta(t,i) =  N_{t_i} / (A_i \cdot F_i ), 
 ~~~~~~\mbox{for every POI }i \mbox{  and every time } t
\end{align}

Focusing at a specific POI, each agent is assigned a dwell time ($w$) when visiting the POI, which is sampled from the SafeGraph database. Recognizing that not all contacts involve an infected agent, we incorporate the fraction of infected agents ($I_f$) out of the total number of agents to adjust individual contact rates. We assume that agents diligently practice social distancing within buildings and are uniformly distributed within them. To calculate the distance between a pair of agents within a building, we utilize the inverse of the density of people within that building ($1/\delta(t,i)$) as the basis for each agent's contact area ($A_c$). Each agent's contact area is represented as a square with a side length denoted as $r$. The value of $r$ is simply calculated by $\sqrt{A_c}$. Consequently, the distance $d$ between a pair of agents is approximated by $r$. The probability of transmission between two agents within the building is denoted as $P_c(d,m)$, where $d$ signifies the distance and $m$ serves as a binary index indicating whether the agents are wearing masks or not ($P_c(d,m)$this information is sourced from Agrawal and Bhardwaj \cite{agrawal2021probability} that the dynamics of airflow and droplets are discussed). $C$ represents the contact rate based on the age groups that are present at the POI (based on the research on the contacts per day for different age groups by Feehan and Mahmud \cite{feehan2021quantifying}. We assumed that the daily contacts during the active hours of the day excluding the night times). It should be noted that $P_{tr}$ is updated by arrival and departure of an agent to the POI.

Lastly, we calculate the probability of infection transmission for a randomly selected pair of agents sharing the same building using the formula:
\begin{align}
P_{tr} = 1 - P_c(d,m)^{C \cdot w \cdot I_f}
\label{eq:p_transmission}
\end{align}

\subsection*{Mobility patterns}

This section is focused on reconstructing mobility patterns, which describe destination types and geographic information extracted from historical data. Agent mobility plays a pivotal role in pandemic dynamics. Analyzing SafeGraph mobility patterns reveals that agent choices regarding destinations exhibit a correlation with the distance to those destinations. Notably, the assumption of equal visit probability for similar types of destinations doesn't hold for certain types, where preference is often given to the closest location – referred to as local trips. Non-local trips correspond to destinations where geographical proximity's impact is minimal, describing long-distance travel and special events. Notably, the COVID-19 pandemic led to a substitution of long-distance trips with local trips, resulting in structural changes within the mobility network while maintaining overall mobility, as demonstrated in Schlosser et al. \cite{schlosser2020covid}.

Our examination of the SafeGraph mobility data underscores the significance of proximity for three distinct types of Points of Interest (POIs): 1) Groceries and department stores, 2) Education services, and 3) Religious organizations. Within the total mobility data analyzed for the year 2021 in the U.S., these three POI types contribute 25\%, 6.8\%, and 1.8\% of trips, respectively. Impressively, over 75\% of the trips are directed towards the five closest destinations. These mobility insights not only aid in the development of a realistic simulation but also hold potential for refining the geographical abstraction of the simulation scope, which will be elaborated upon in subsequent discussions.

\subsection*{Tessellation types}

Tessellation plays a pivotal role in facilitating agent-based simulations by dividing the simulation space into discrete cells or polygons, effectively organizing and governing agents' interactions. This spatial partitioning not only streamlines agent tracking and interaction management but also enhances the simulation's scalability to encompass larger systems. Commonly employed tessellation techniques encompass regular grids, Voronoi diagrams (VDs), Delaunay triangulation, among others. Each tessellation cell is furnished with a mobility schedule, outlining a roster of Points of Interest (POIs) and corresponding visit probabilities for distinct demographic groups. The connection between mobility from SafeGraph and demographic groups is not explicitly available. We used the needs of agents from Mahmood et al. \cite{mahmood2020facs} research. We aggregated their needs matrix into three age categories and we connected the needs to NAICS codes. Mobility schedule of a tessellation cell is the list of POIs that historically visited by the tessellation cell based on SafeGraph. We store the frequencies related to the list of POIs for each tessellation cell and we convert the frequencies to probability distribution before starting the simulation. By storing the visit frequencies for each tessellation cell, it is possible to split and merge schedules based on new constructed tessellation.

In exploring various tessellations, it becomes imperative to grasp the underlying significance of tessellation in the context of large-scale agent-based simulations. Primarily, the adoption of tessellation is necessitated by the intrinsic nature of general-purpose mobility, which involves trajectories bridging one area (tessellation cell) to another. While certain mobility datasets pertain to specific point-to-point journeys, as observed in public transportation data, the preservation of individual privacy is inherent due to the shared nature of such modes. Nonetheless, for comprehensive mobility datasets, it is paramount to avoid pinpointing precise origins and destinations of trajectories. A second rationale for embracing tessellation relates to the intricate interplay of mobility. When the mobility patterns within a tessellation cell exhibit a higher degree of uniformity, it becomes reasonable to describe the cell using more abstract definitions. Depending on the modeling approach, certain agent activities might be confined to cells that represent communities or neighborhoods, thus contributing to more efficient simulations. Thirdly, tessellation cells serve as a valuable framework for implementing control policies, and a reduced number of cells is favorable while ensuring that the requirements of each cell are met. Mobility is fundamentally driven by individual needs. When a tessellation cell adequately caters to the majority of an individual's needs within their residing area, it leads to diminished mobility between distinct cells. Moreover, the choice of tessellation type can also optimize memory resources.

Prior to discussing technical details, we would like to make a general remark. Our study primarily focuses on enhancing the computational efficiency and scalability of infectious disease modeling through innovative approaches such as tessellation and the use of super-agents, we acknowledge the potential implications of our work for public health strategies. Specifically, the introduction of tessellation cells within our model offers a structured framework that could be instrumental in the design and implementation of geographically informed control policies. Although our current research does not directly investigate specific control policies, the geographical delineation provided by tessellation cells facilitates a more nuanced understanding of disease spread within distinct regions of a city. This, in turn, could aid policymakers and public health officials in devising targeted interventions based on the spatial dynamics of an outbreak. By defining regions within a city with precision, our model lays the groundwork for future studies to explore how control policies can be optimally applied in a real-world context, thereby contributing to the broader goal of improving disease modeling and public health response strategies.

\subsubsection{Voronoi diagram on network}

Employing a Voronoi diagram to partition a city into a tessellation proves to be a valuable and enlightening method for urban planning and analysis. The Voronoi diagram, a geometric construct, divides space into distinct regions determined by their proximity to specified points of interest (POIs). In the context of urban division, these points serve as pivotal locations like public amenities, transportation hubs, or prominent landmarks. Each region within the Voronoi diagram corresponds to the space closest to a specific seed point compared to all others.

Leveraging Voronoi tessellation in city planning empowers urban designers and policymakers with invaluable insights into the spatial arrangement of facilities and services. This approach proves particularly useful in delineating service coverage zones for essential institutions such as hospitals, schools, police stations, and fire departments. Strategic placement of seed points at optimal positions guarantees equitable access to these critical services across each region, enhancing the efficiency and effectiveness of urban infrastructure as a whole.

Moreover, the Voronoi diagram (VD) plays a pivotal role in comprehending the spatial intricacies of an urban environment. It effectively demarcates neighborhoods and establishes their borders through proximity to diverse facilities, enabling urban planners to scrutinize the influence of amenities on the city's structural composition. This insight proves indispensable for informed deliberations concerning upcoming urban expansions and zoning directives.

An additional benefit of employing Voronoi tessellation lies in its inherent scalability. As a city undergoes expansion and transformation, the integration of new facilities can be seamlessly accomplished by incorporating them as seeds. Consequently, the Voronoi diagram can be readily recalibrated to incorporate these modifications. This inherent flexibility renders it an invaluable resource for accommodating the complexities of dynamic urban planning situations.

Utilizing Voronoi diagram-based tessellation represents a potent methodology for partitioning a city into distinct zones predicated on their proximity to vital amenities and services. This approach not only furnishes invaluable perspectives into urban dynamics but also facilitates the crafting of resourceful, just, and flexible urban environments. By harnessing this technique, urban planners are empowered to make well-informed choices that elevate residents' quality of life and amplify the overall efficacy of the urban setting.

The Voronoi diagrams (VDs) are constructed based on commuting time along the street network, taking into account route types and average traffic load. To create VDs using commuting time, a method akin to Dijkstra's search algorithm is employed. Rather than searching for destinations, this method calculates travel times from the multiple sources. Each source leaves traces on the map while calculating travel times, and the node in the graph is won by the source  with minimum commuting time. The source code for the implemented Agent-Based Model (ABM) is accessible online\footnote{https://bitbucket.org/pandemic-ames/aicsimuation}. 
Efficiently implementing this data-rich Agent-Based Model (ABM) demands the utilization of streamlined data structures and precomputed queries. The case studies conducted in this research, involving preprocessed geographical data, are readily accessible within the online repository. The construction of Voronoi Diagrams (VDs) can be performed in parallel. Because each node on the OpenStreetMap (OSM) map is claimed by a single VD cell, synchronization is unnecessary throughout this process. Consequently, the order in which Voronoi Diagram cells assert ownership over nodes on the map is inconsequential. The definitive owner of each node is ultimately determined by comparing the traces left by each VD cell. Numerous opportunities exist to leverage parallel processing for the construction of VDs. Two implemented approaches include: 1) parallel construction based on Voronoi Diagram types, such as groceries, schools, and religious locations; and 2) parallel construction for each individual Voronoi Diagram cell. VDs preserve the CBG boundaries if a CBG is cut by a VD cell. If two ore more CBGs are fall inside a VD cell, they are merged together. When a CBG is cut by a VD cell, two cells are generated for VD tessellation and the schedules related to each generated cell is based on the area of the cut. For instance, if a VD boundary cuts a CBG cell by 10\% and 90\% area into two cells $C_1$ and $C_2$, the schedules for $C_1$ will contain 10\% of the POI visits' frequencies. Additionally, 10\% of the demographic groups are assigned to $C_1$.

VDs are constructed on street network as vectors. To convert the VDs to land area, we map the VD vectors to a high resolution pixel map and we grow pixels related to each VD cell to the adjacent unallocated pixels. Figure \ref{fig:VD_Seattle_withSubfigures} shows a visualization of mapping VD street vectors to a pixel map.

\subsubsection{Euclidean clustering}

To underscore the significance of network-based commuting time, we turn to  the concept of Euclidean clustering, where each street node functions as a data point and a tessellation cell emerges from clusters of street nodes sharing proximity. In this approach, conventional clustering algorithms like the K-means algorithm can be applied. Unlike the Voronoi Diagram (VD) tessellation, in this scenario, cluster centers are established through the K-means algorithm, with the number of clusters determined by the number of cells generated by the VD tessellation. The core focus of this tessellation construction is the amalgamation of geographical street nodes.

K-means clustering, a popular unsupervised machine learning technique, serves the purpose of data clustering and pattern recognition. The algorithm's objective is to partition a given dataset into $K$ distinct clusters, where each data point is allocated to the cluster with the nearest mean (centroid). The process entails initializing $K$ centroids randomly or through heuristics, followed by an iterative sequence of assignment and update steps. During each iteration, data points are assigned to the closest centroid based on their distance, often measured by Euclidean distance, from the centroids. Subsequent to the assignment phase, the centroids are recalculated by determining the mean of the data points within each cluster. The algorithm iterates until centroids exhibit minimal change or a predefined iteration limit is reached. K-means is computationally efficient and effective for sizable datasets, rendering it applicable across domains such as image segmentation, customer categorization, and anomaly detection.

Though K-means is extensively employed and relatively straightforward to implement, it does carry certain limitations. Notably, it's sensitive to the initial positioning of centroids, which can result in suboptimal clustering outcomes or convergence to disparate local optima. To counteract this concern, the algorithm is often executed multiple times with varying initializations, and the most favorable clustering result is selected based on a pre-established evaluation metric \cite{arthur2007k}.

\subsection*{Reduced number of agents}

Agents are in the cornerstone of the pandemic dynamic simulation mechanism. The quantity of agents in an agent-based simulation stands as a critical determinant in shaping the model's realism, dynamics, and overall efficacy. Considering that agents directly impact computational time, reducing their numbers can yield substantial performance gains while preserving simulation dynamics. Conversely, a larger agent population permits a more nuanced emulation of actual systems, fostering better approximations of individual behaviors, interactions, and emergent patterns. By integrating a robust agent contingent, the simulation gains the versatility to simulate diverse scenarios and accommodate variations in individual decision-making processes, amplifying both predictive and explanatory potentials. Moreover, an ample agent count facilitates exploration of rare or unforeseen events that might materialize only under specific agent densities, thus unveiling deeper insights into system dynamics and potentially identifying pivotal tipping points. Analyzing the influence of agent numbers on simulated pandemic patterns offers valuable insights. Evidently, sparse agent representation can lead to pronounced deviations from authentic mobility trajectories.

Modeling the pandemic should strive to minimize the simulation's sensitivity to the agent count, ensuring robustness regardless of the number of agents employed. For instance, an agent's at-home activities should be contingent on family size, remaining independent of the total agent count within the study scope. This rationale should be thoughtfully extended to all aspects of the ABM, promoting a design that fosters stability and meaningful results regardless of the specific agent population.

Given that the infection model relies on both agent-agent and agent-environment interactions, altering the agent count can lead to shifts in pandemic trends. To mitigate this dependency on agent count, two pivotal modifications should be made to the proposed ABM. Firstly, maintaining consistent congestion estimations within Points of Interest (POIs) is imperative, relying on SafeGraph data rather than agents' visits for accuracy. Secondly, a novel agent type, termed "super-agent" (SA), is introduced to address this concern. SAs act as aggregators, representing groups of individual agents within a single entity. To elaborate, at a POI, infection transmission relies on the ratio of infected to healthy agents, which becomes less reliable as agent count decreases. This discrepancy emerges due to the reduction in the numerator and denominator of the ratio, stemming from a smaller number of agents present at the POI simultaneously. To rectify this, the SA concept is proposed. A SA embodies the collective medical status of a group of agents, maintaining a single physical representation while holding distinct medical statuses. For instance, with agent count reduced to one-fifth of the population, each SA would encapsulate five unique medical statuses. This signifies that multiple agents share a singular physical presence, yet exhibit diverse medical characteristics. Notably, each SA corresponds to agents sharing common demographic attributes, effectively encapsulating the behavior of similar agents throughout the simulation.

A SA represents a group of agents for mobility activities. For instance when a SA moves from a CBG to a POI, it means that a group of agents made the trip together. However, inside a POI and during the activities, the agents that are represented by SAs act similar to a group of real agents. The reason that SAs can act similar to a group of real agents in the mentioned cases, is that the activities and the disease transmission model only require the number of agents and the number of infected agents. This means that if there is only one SA at a POI, disease transmission can still occur by the agents represented within a SA. The accuracy of infection fraction i.e. $I_f$ in equation \ref{eq:p_transmission} is sustained by SA.

Employing SAs ensures the preservation of both mobility and infection dynamics to the greatest extent possible. Nevertheless, in scenarios where the agent count is dramatically decreased, the availability of sufficient SAs becomes a crucial concern. This situation could potentially lead to inadequate SA representation within the simulation. For example, a tessellation cell might lack the requisite number of SAs to adequately represent a specific demographic group. In essence, maintaining an appropriate balance between agent count and the corresponding SA population is essential to uphold the accuracy and validity of the simulation's results.

\section*{Experiments: Setting}

The urban structure of different cities can vary significantly due to factors such as population density, geography, and culture. Assuming that total population is a crucial determinant, we conduct three case studies: Santa Fe, NM; Seattle, WA; and Chicago, IL. These cities represent cities with low, medium, and high population, respectively. In order to explore the impact of tessellation type and the number of agents, we propose three metrics centered around the concept of the Most Visited Point of Interest (MVPOI): (a) The Number of Visits (NOV) to the MVPOI in each city over the course of a week; (b) The Average Visit Duration (AVD) at the MVPOI; (c) The Probability of Co-Visiting (PCV) the MVPOI by two agents during the same time period. 

Let us offer some clarifications regarding our selection of Most Visited Point of Interest (MVPOI) for both validating tessellations and establishing the extent of population coarse-graining (i.e., the number of super-agents). First, it's essential to note that MVPOIs differ for each residential cell. Second, MVPOIs vary across distinct types of Points of Interest (POIs), such as groceries and medical facilities, within a given residential cell. Third, our exclusive focus on MVPOIs warrants an explanation. This choice is motivated by the fact that less frequently visited POIs might exhibit a significantly lower count of visits, thereby rendering the collection of reliable statistics unfeasible. As a result, we concentrate on MVPOIs as they provide a robust basis for analysis and interpretation.

We select $CBG_1$ and $CBG_2$ by the highest source of visits to MVPOI. Assuming that an agent from $CBG_1$ ($a_1$) visits MVPOI, if an agent from $CBG_2$ ($a_2$) is not present at MVPOI a counter related to trip without a co-visit is incremented. If $a_1$ and $a_2$ are present at MVPOI and they stay for more than five minutes together the counter for co-visit is incremented. PVC is the fraction of co-visit counter and total visits from $CBG_1$ and $CBG_2$.

To further accentuate the influence of tessellation type, we also introduce a novel tessellation created by merging neighboring Census Block Groups (CBGs), referred to as Randomly Merged CBG (RMCBG) tessellation. The tessellations examined in our experiments are summarized below:
\begin{itemize}
\item $\textit{VD}_r$: VD constructed based on Points of Interest (POIs) related to retail services, educational services, and religious organizations, categorized by commuting time. Here, the number of generated cells is fewer than the number of CBGs.

\item $K$-$means_r$: Involves clustering streets, treated as nodes, using Euclidean distance. The parameter $K$ is set to the number of cells generated within $\textit{VD}_r$.

\item $RMCBG$: Constructed by (random) merging adjacent CBG cells to match the cell count generated in $\textit{VD}_r$.

\item $CBG$: This tessellation is based on census data.

\item $VD_s$: Based on VD of all POIs sorted by visit frequency. In this case, the number of cells matches the number of CBG cells.

\item $K$-$means_s$: Similar to $K$-$means_r$, but with $K$ set to the number of CBG cells.

\item $CBG\textit{VD}$: A hybrid tessellation overlaying VD tessellation and CBG tessellation, resulting in breaking original cells into a larger number of smaller cells.

\item $VD_i$: Similar to $VD_s$, but with the number of cells determined by the CBGVD tessellation.

\item $K$-$means_i$: Similar to $K$-$means_r$, but with $K$ equal to the number of cells in the CBGVD tessellation.

\end{itemize}

The agents are placed on the map according to their home locations, determined by a combination of population density and street types (treated as nodes in a geo-graph). We design this placement to prioritize residential streets, making it highly likely for agents to be positioned there, while virtually eliminating the probability of agents being placed on highways. We note that the number of Census Block Groups (CBGs) is 62 in Santa Fe, 485 in Seattle, and 2180 in Chicago.

We introduce a fundamental -- ground truth -- point of reference for our experiments, denoted as the No Tessellation (NT) approach. In the NT method, we leverage statistical data extracted from SafeGraph and commuting times to Points of Interest (POIs). However, unlike considering mobility from Census Block Group (CBG) to POI, NT takes into account mobility originating from precise locations to the respective POIs. Subsequently, we develop tessellations based on a map featuring an array of potential trajectories. Naturally, we anticipate that tessellations exhibiting more uniform mobility patterns will offer distinct advantages.

In each experiment, employing a specific tessellation variant alongside a reduced number of agents, we yield metrics including Number of Visits (NOV), Average Visit Duration (AVD), and Probability of Co-Visiting (PCV), all comparable with the NT baseline. We explore three distinct sets of tessellations categorized by the number of generated cells. The first group maintains an equivalent (the same) number of cells ($s$) as the original Census Block Groups (CBGs), while the second group produces fewer (reduced number of) cells ($r$), and the third group generates an increased quantity of cells ($i$). For the tessellations with fewer cells, the count is determined by the Voronoi Diagrams (VDs) of the POIs. In contrast, tessellations generating more cells than the base case result from an overlay of VDs and CBGs.

To ensure comprehensive insights into the variations across experiments, we conduct each experimental setup several times, thereby providing a robust depiction of the spectrum of changes compared to other trials. We observed that the average of the performance measures and runtime converges after 72 replications. We used change point detection on mean and standard deviation and we used the largest number of replicates that shows convergence to ensure that all measures are converged \cite{killick2012optimal}.

We compare our proposed model, coined AIC, with FACS-CHARM \cite{anagnostou2022facs}, a state-of-the-art agent-based model that aligns closely with our approach. FACS-CHARM builds upon the FACS pandemic simulator \cite{mahmood2020facs}, offering enhanced details within hospital simulations, including discrete simulations of wards and intensive care units. As one of the most recent advancements in ABMs, it incorporates the geo-spatial features from its predecessor, the FACS simulator. The simulation integrates building data extracted from OSM, and agent mobility is determined by individual needs and distances, positioning FACS as one of the most geographically precise simulators available. This level of detail and alignment with our model's aims is why we specifically chose to compare AIC with FACS-CHARM, despite the plethora of other models in literature dedicated to simulating pandemic spread.

While FACS-CHARM utilizes a relatively straightforward infection model and simulates agents on a daily basis, our model, AIC, offers a minute-by-minute simulation, capturing distinct actions each hour. Our system preprocesses agent groupings—referring to daily tasks and transportation—on an hourly basis, in contrast to FACS-CHARM's daily processing routine. To optimize speed, our model employs large binary files for data preprocessing storage, whereas FACS-CHARM utilizes comma-separated text data, a method generally slower. For an equitable runtime comparison, we exclude data preprocessing and loading times for both models.

The mobility of agents in FACS-CHARM reflects their individual needs, whereas in AIC, mobility is determined based on probabilities extracted from commuting times and SafeGraph data. To conduct a practical comparison, we utilize data from FACS-CHARM for the Brent neighborhood of London and simulate Lexington, Kentucky, due to its similar population size. Our focus is on assessing the runtime and infection trends over a 10-day period for both simulators. Given that FACS-CHARM's simulation for Brent operates independently of the rest of London, it parallels our study of Lexington, which is distinctly separated from neighboring urban areas.

\section*{Experiments: Results}

\begin{table}[h!]
\centering
\begin{tabular}{l|c|c|c}
 & Chicago & Seattle & Santa Fe \\
 \hline
 Reduced cells  & 493 & 242 & 52 \\
 Cells in CBG & 2179 & 485 & 62 \\
 Increased cells & 5051 & 1032 & 373 \\
 Retailing services & 74 & 56 & 13 \\
 Educational services & 161 & 81 & 21 \\
 Religious organizations & 49 & 32 & 14 \\
 Population & 2,722,000 & 679,000 & 86,400 \\

\end{tabular}
\caption{Extracted information from analyzing the case studies.}
\label{table:CSA}
\end{table}

\begin{figure}[h!]
\centering
\captionsetup{justification=centering}
\begin{subfigure}[b]{0.24\textwidth}
  \centering
  \includegraphics[width=.95\linewidth]{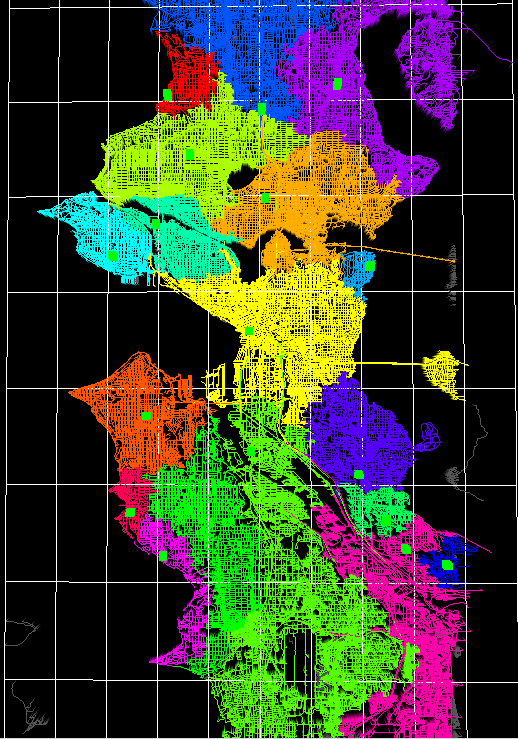}  
 \caption{}
  \label{fig:VD_seattle_shop_school_sub_first}
\end{subfigure}%
\begin{subfigure}[b]{0.24\textwidth}
  \centering
  \includegraphics[width=.95\linewidth]{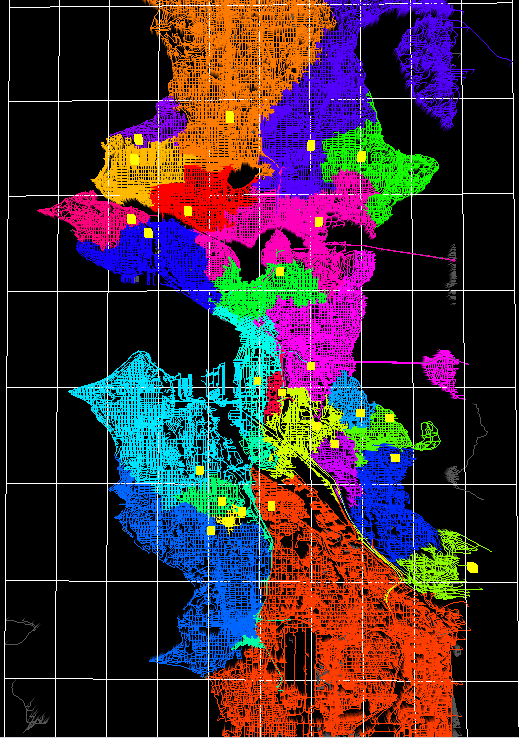}  
 \caption{}
  \label{fig:VD_seattle_shop_school_sub_second}
\end{subfigure}
\begin{subfigure}[b]{0.24\textwidth}
  \centering
  \includegraphics[width=.96\linewidth]{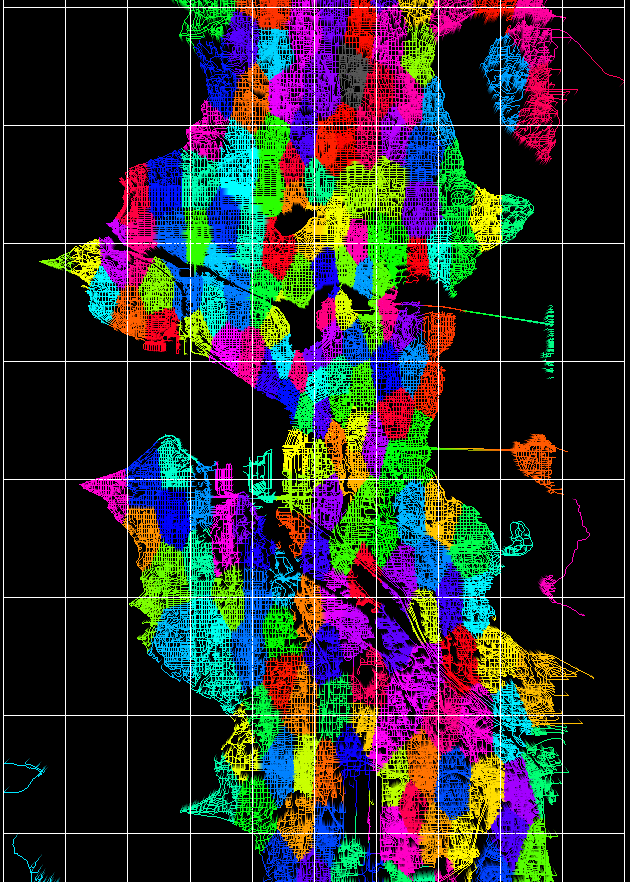}  
 \caption{}
  \label{fig:VD_seattle_kmeans}
\end{subfigure}
\caption{ Seattle map. {\bf (a):} Voronoi Diagram depicting tessellation for shopping centers; {\bf (b):} Voronoi Diagram showcasing tessellation for schools;
{\bf (c):} K-means clustering generating 242 distinct cells.
\label{fig:VD_Seattle_withSubfigures}}
\end{figure}

The counts of cells resulting from the reduced and increased tessellations are presented in Table \ref{table:CSA} across our test cities. Notably, in smaller cities, these tessellations tend to approximate the number of Census Block Group (CBG) cells more closely. On the other hand, when considering the three cities, the cell counts generated through the Points of Interest (POIs) are fewer than the CBG cell numbers. Illustrating the Voronoi Diagram (VD) cells based on commuting times, Figure \ref{fig:VD_Seattle_withSubfigures} provides visual insight. We observe that VD cells can exhibit intricate boundaries, due to the fact that the VDs, which are generally constructed upon the street network, can also be defined and shaped by prominent landmarks.

\begin{figure}[h!]
    \captionsetup{justification=centering}
    \centering
    \includegraphics[scale=0.31]{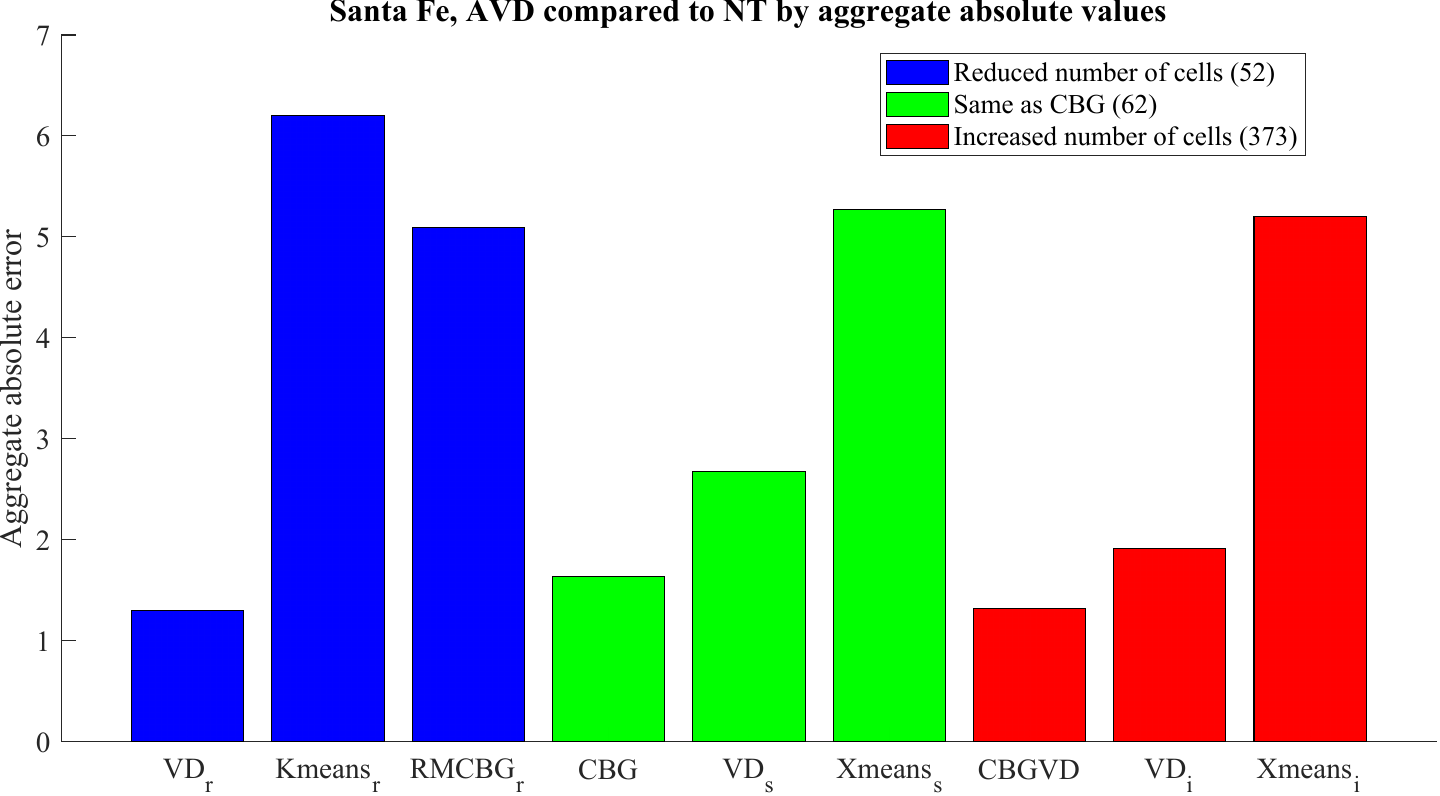}
    \caption{All tessellations for Santa Fe case study showing the aggregate absolute error for the Average Visit Duration (AVD) measure. Aggregate absolute error is the average absolute error over all agent reduction scenarios.}
    \label{fig:SantaFeAVDAll}
\end{figure}

We study performance of the AIC (Agent-In-Cell) model over a week, initializing agent placement on the map based on the historical COVID-19 cases sourced from the John Hopkins’ dataset \cite{dong2020interactive}. Aggregating performance outcomes across various agent reduction scenarios, Figure \ref{fig:SantaFeAVDAll} summarizes the effectiveness of all tessellation strategies. Aggregate absolute error is calculated by averaging absolute error of AVD between no-tessellation and different tessellation types over different agent reduction scenarios i.e. 75\%, 50\%, 25\%, and 10\%. Notably, tessellations relying on K-means Euclidean clustering exhibit suboptimal performance.

\begin{figure*}[h!]
    \captionsetup{justification=centering}
    \centering
    \includegraphics[width=0.9\textwidth]{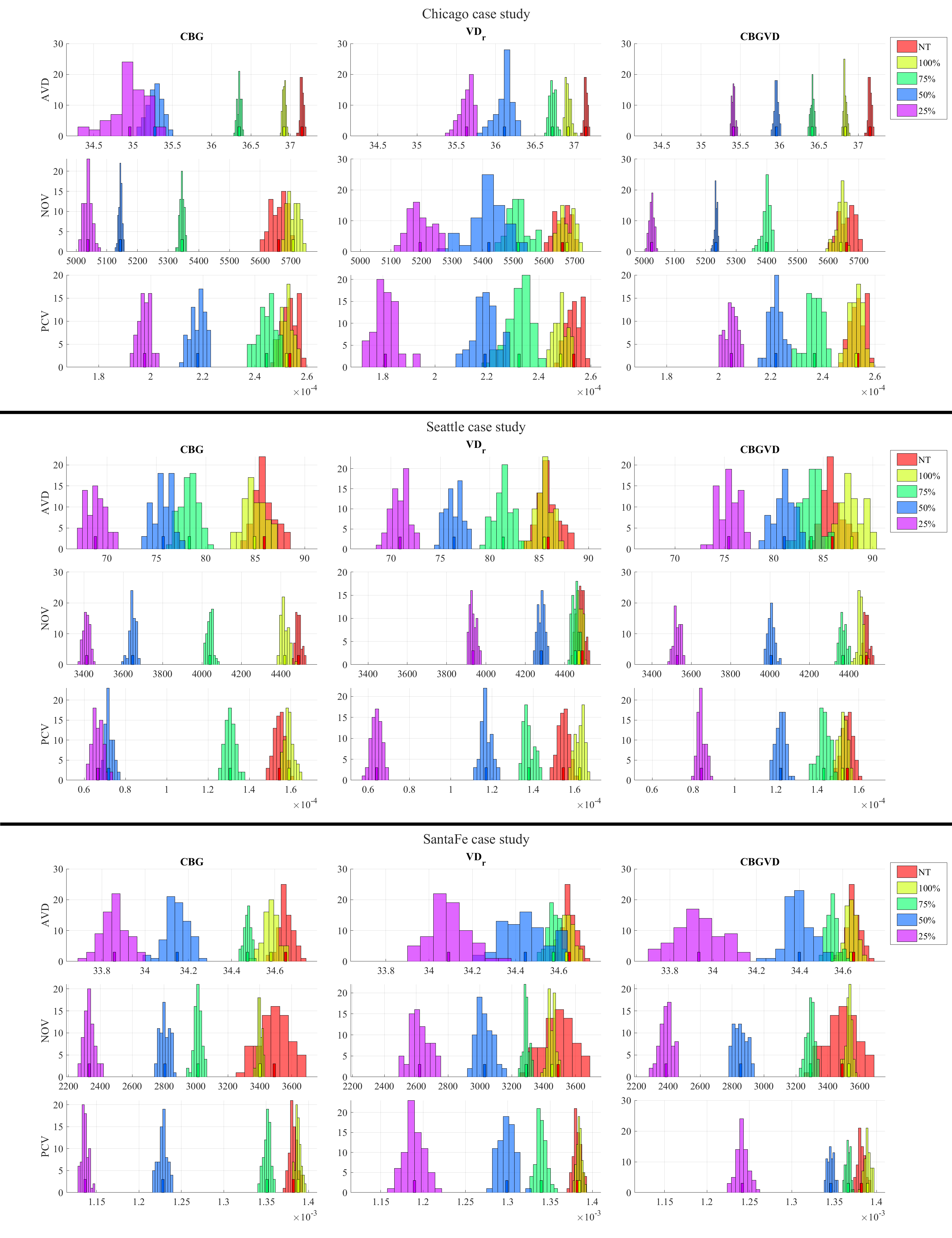}
    \caption{Histograms (binned probability distributions) illustrating the impact of three primary tessellations and a reduced number of agents, quantified through three distinct measures. (See text for details.)}
    \label{fig:allGoodTessellations}
\end{figure*}

To gain further insight, we closely examine the three most successful tessellations (showing in Figure \ref{fig:SantaFeAVDAll} least aggregate absolute error) in Figure \ref{fig:allGoodTessellations} -- $CBG$, $\textit{VD}_r$ and $CBG\textit{VD}$ -- scrutinizing their behavior across three key measures and their response to reductions in the number of agents. The Figure (most comprehensive in the manuscript) illustrates the Average Visit Duration (AVD), Number of Visits (NOV), and Probability of Co-Visiting (PCV) metrics across a spectrum of city sizes --Santa Fe, NM (small), Seattle, WA (medium), and Chicago, IL (large). Notably, all the recorded data pertains to the Most Visited Points of Interest (MVPOI) and originates from the SafeGraph consortium database \cite{SafeGraphDataConsortium}. It is also important to emphasize that Figure \ref{fig:allGoodTessellations} presents histograms (binned probability distributions) where varying color codes correspond to distinct levels of coarse-graining, representing different numbers of SAs. The reference point for comparison is the "red" No Tessellation (NT), serving as the ground truth. Therefore, a more pronounced deviation from the NT histogram signifies a greater divergence from the precise representation.

The following insights are gleaned from a comprehensive analysis of the results reported in Figure \ref{fig:allGoodTessellations}:
\begin{enumerate}

\item \underline{Impact of Agent Count:} A discernible decline is noted with a reduction in the number of agents. This underscores the significance of our methodology in determining an optimal level of coarsening—essential for striking a balance between desired quality and result assurance.

\item \underline{Depletion in PCV:} The most substantial degradation is observed in the Probability of Co-Visiting (PCV) metric. This observation is logical given the inherent complexity of this feature among the three under consideration.

\item \underline{Comparable Performance:} The three winning strategies -- VD, CBG, and CBGVD -- demonstrate approximately equivalent performance. Notably, CBGVD outperforms the rest in modeling the most intricate feature (PCV) and also fares the best in the case of Average Visit Duration (AVD).

\item \underline{VD for Visit Count:} The VD strategy emerges as the most effective in capturing the modeling of visit counts (NOV).

\item \underline{City-Specific Intricacies:} The characterization of certain cities exhibits nuances that transcend mere population size. Notably, the Average Duration of Visits in Chicago (the largest city) and Santa Fe (the smallest) reveals striking similarity, while Seattle (a medium-sized city) stands out with a nearly twofold longer average duration.

\item \textbf{Variation in Simulations:} With an increase in the number of cells in a tessellation, the variation from repeated simulations diminishes. Furthermore, the overlap extent across different levels of agent reduction notably diminishes as the number of cells rises.

\end{enumerate}

These observations collectively provide valuable insights into the behavior and efficiency of different tessellation strategies across various dimensions of our analysis.

Fig.~\ref{fig:allGoodTessellations} illustrates the synergistic effects of tessellation and super-agents (SA) on our simulation outcomes. To distinguish the impact of SAs, we conducted simulations for the city of Seattle without employing SAs, thereby reducing the number of agents without the explicit use of SAs. The contribution of SAs to the simulation metrics is quantified by comparing outcomes with and without the use of SAs. For instance, if the Average Visit Duration (AVD) without tessellation is $\mathrm{AVD}_{\mathrm{nt}}$, the AVD with Voronoi Diagram (VD) tessellation using SAs is $\mathrm{AVD}_{\mathrm{VD_{s}}}$, and the AVD for VD tessellation without SAs is $\mathrm{AVD}_{\mathrm{VD_{ns}}}$, then the contribution of the SA is calculated by:
\begin{equation*}
    \tau=\frac{\mathrm{AVD}_{\mathrm{VD_{ns}}}-\mathrm{AVD}_{\mathrm{VD_{s}}}}{\mathrm{AVD}_{\mathrm{nt}}-\mathrm{AVD}_{\mathrm{VD_{ns}}}}
\end{equation*}

\begin{figure}[h!]
    \captionsetup{justification=centering}
    \centering
    \includegraphics[scale=0.65]{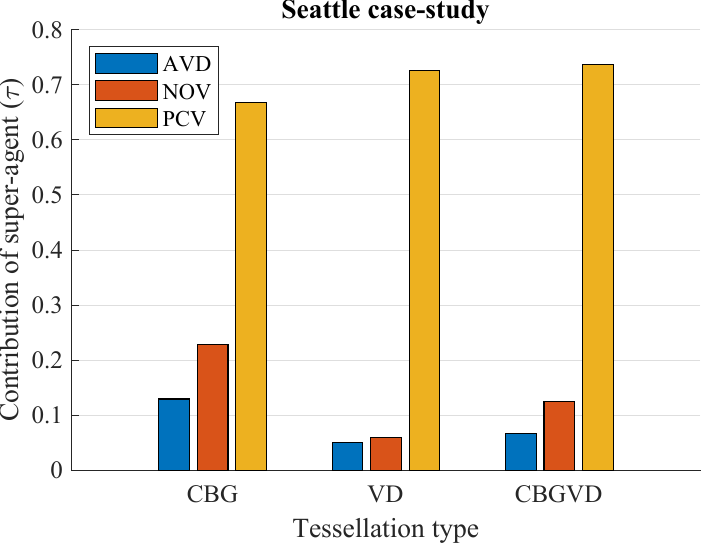}
    \caption{Evaluating the Impact of Super-Agents Across Various Tessellation Strategies on Simulation Outcomes in the Seattle Case Study. This figure demonstrates how the incorporation of super-agents with different tessellation methods compares to a baseline scenario without tessellation, emphasizing the role of super-agents in improving the precision of model outcomes.}
    \label{fig:superAgentContribution}
\end{figure}

This approach is also applied to quantify the effects of SAs on the Number of Visits (NOV) and the Peak Congestion Volume (PCV) metrics. Fig.~\ref{fig:superAgentContribution} presents the contribution of SAs to these simulation outcomes under various tessellation strategies for the Seattle case study, revealing that SAs significantly improve the PCV metric, whereas their impact on AVD and NOV is more marginal. It is important to note that SAs enhance the accuracy of disease transmission dynamics; an infected agent's need for hospitalization or self-quarantine indirectly influences mobility patterns. From Figure \ref{fig:superAgentContribution}, it is evident that SAs substantially affect the PCV measure, with a marginal effect on AVD and NOV.

\begin{figure}[h!]
    \captionsetup{justification=centering}
    \centering
    \includegraphics[scale=0.65]{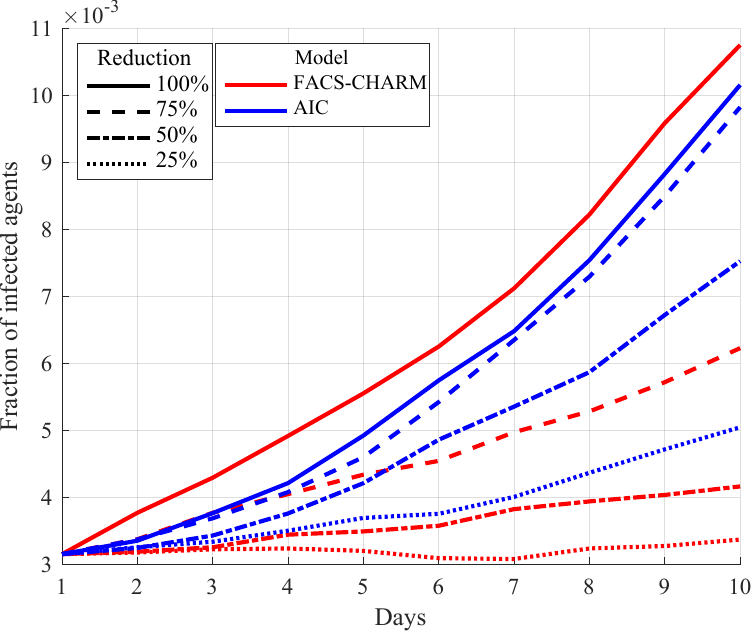}
    \caption{Comparison of the fraction of infected agents over a 10-day simulation between AIC and FACS-CHARM with std error bars.  AIC is evaluated on Lexington, Kentucky, and FACS-CHARM is tested on Brent, London. We observe that the AIC, employing super-agents during agent reduction, effectively maintains the infection spread better with fewer agents.}
    \label{fig:AIC_FACSCHARM_REDUCTION}
\end{figure}

Our AIC model excels in replicating pandemic dynamics with fewer agents compared to the FACS-CHARM model \cite{anagnostou2022facs}. As illustrated in Figure \ref{fig:AIC_FACSCHARM_REDUCTION}, the AIC model's behavior closely mirrors that of a comprehensive simulation (100\% of the total agent count), using only 75\%. Notably, the AIC model exhibits a more pronounced decline in disease spread as the number of agents decreases, unlike the FACS-CHARM simulator. This efficiency is attributed to the use of super-agents and Voronoi Diagram tessellation.

\begin{figure}[h!]
    \captionsetup{justification=centering}
    \centering
    \includegraphics[scale=0.65]{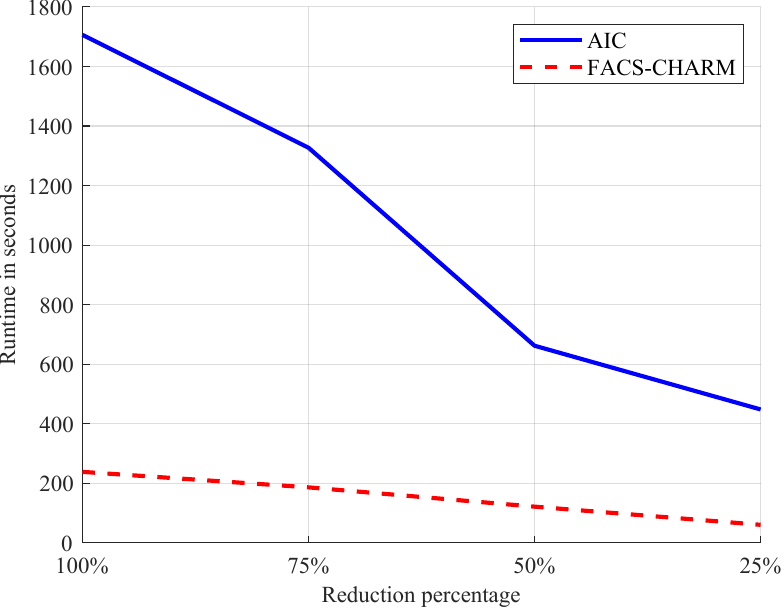}
    \caption{Comparison of runtime between AIC and FACS-CHARM simulators across varying agent counts, highlighting the impact of super-agents in AIC for agent reduction efficiency. AIC evaluations are conducted on Lexington, Kentucky using 8 processing threads without tessellation, while FACS-CHARM assessments are done on Brent, London. Notably, although AIC exhibits a longer runtime compared to FACS-CHARM, it demonstrates greater resilience to changes in agent numbers. This is due to AIC's more frequent rescheduling and regrouping of agents (hourly vs. daily in FACS), its simulation of detailed actions every minute within each hour (unlike FACS, which does not simulate actions and operates only on a day-scale), and the minute-by-minute simulation of building congestion and pathogen spread, leveraging SafeGraph's minute-scale data on the duration of stays at Points of Interest (POIs). These features contribute to AIC's comprehensive simulation capabilities, albeit with a trade-off in computational speed.}
    \label{fig:AIC_FACSCHARM_REDUCTION_RUNTIME}
\end{figure}

In terms of runtime, we conducted comparative tests on an AMD Ryzen 7 1700X PC. Fig.~\ref{fig:AIC_FACSCHARM_REDUCTION_RUNTIME} displays the runtime for both simulators relative to the number of agents. We observed that the FACS-CHARM simulation runs more quickly than AIC, for three main reasons: (a) FACS reschedules and regroups agents once per day, while AIC does this hourly to align with their tasks and locations, enhancing the accuracy of agent contacts. (b) FACS operates solely on a daily timescale without simulating actions, whereas AIC conducts minute-by-minute checks for hourly-specific actions, thereby more accurately simulating the pandemic's reproduction number. (c) AIC simulates building congestion and pathogen persistence every minute, providing a more detailed transmission simulation, especially in scenarios involving brief visits by infected individuals.

\begin{figure}[h!]
    \captionsetup{justification=centering}
    \centering
    \includegraphics[scale=0.65]{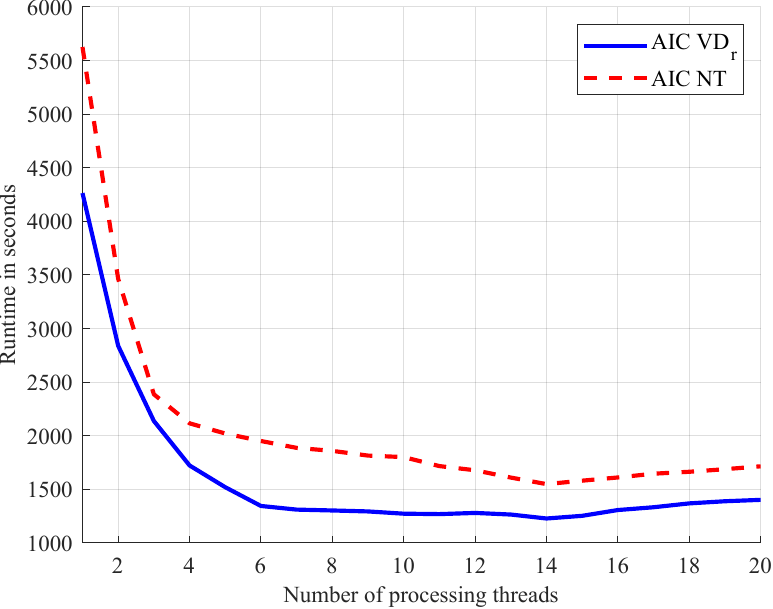}
    \caption{AIC runtime as a function of processing thread count for both $NT$ and $\textit{VD}_r$ tessellation methods.
}
    \label{fig:AIC_THREAD_RUNTIME}
\end{figure}

AIC's use of super-agents helps maintain accurate pandemic dynamics, and it is capable of parallelizing certain aspects of the simulation. While agent behaviors are assessed in parallel, the simulator undergoes a regrouping and synchronization process every hour. It is essential for all processing threads to align during each synchronization, limiting the extent of feasible parallelization. Figure \ref{fig:AIC_THREAD_RUNTIME} demonstrates the impact of adding more processing threads to the AIC simulation. Beyond the addition of 14 threads, there is an unexpected increase in runtime, likely due to the heightened synchronization demands among threads, which hampers performance gains. Furthermore, employing a simplified tessellation approach results in quicker runtimes compared to scenarios without tessellation.

Moreover, the choice of tessellation type can significantly optimize memory resources in the simulation of large-scale urban environments. To substantiate this claim and assess the computational efficiency of our simulations, we meticulously measured memory usage across various tessellation types while simulating different numbers of agents within the AIC framework, which operates atop the Java Virtual Machine (JVM). This empirical analysis, detailed in Table \ref{table:memory}, not only demonstrates a direct correlation between the tessellation strategy and the required memory but also highlights the memory efficiencies achieved through deploying fewer agents.

Given the unique challenge of accurately gauging memory consumption on the JVM, we enhanced our measurement precision by initiating a garbage collection process during periods of peak agent activity. Utilizing the VisualVM tool for monitoring memory consumption and managing garbage collection explicitly allowed us to derive insights into the impact of agent count on simulation resource demands. The findings, as presented in Table \ref{table:memory}, reveal that carefully selecting the tessellation type contributes to significant memory savings, thereby validating our approach to optimizing memory use through strategic tessellation choices and the judicious management of agent simulations.


\begin{table}[!t]
\centering
\captionsetup{justification=centering}
\begin{tabular}{ r|l l l l }
  & \multicolumn{4}{c}{Tessellation types}\\
  Available agent percentage & NT & CBG & VD & CBGVD \\
  \hline
  100\% & 20.08 & 20.64 & 20.27 & 20.98\\
  75\% & 15.84& 16.45 & 16.24 & 16.47\\
  50\% & 11.15 & 11.57 & 11.36 & 11.58\\
  25\% & 6.77 & 7.04 & 6.91 & 7.06\\
  10\% & 4.12 & 4.25 & 4.11 & 4.27\\
 \end{tabular}
 \caption{
 Memory Requirements for Simulating Seattle City at GB Scale. The table illustrates the varying memory needs corresponding to different simulation scenarios, each represented by a row, where a specific percentage of agents relative to the total population of Seattle city is actively simulated.
 }
\label{table:memory}
\end{table}

\section*{Conclusion and Path Forward}

This research paper introduces the Agent-In-Cell (AIC) model, an innovative iteration of the Agent-Based Model (ABM). The AIC model uniquely integrates realistic mobility with all actions represented within specific geographical regions, thereby enhancing the simulation's fidelity to real-world scenarios. We depict infection transmission through a combination of detailed actions, person-to-person contacts, and the persistence of pathogens in the environment. Our proposed model introduces various computational strategies to optimize these detailed simulations, striking a balance between accuracy and efficiency.

Notably, the AIC framework retains the dynamism of a comprehensive simulation while employing a reduced number of agents. We have demonstrated that meticulous urban tessellation offers dual benefits: it not only minimizes the count of tessellation cells for memory efficiency but also leads to a reduction in agent numbers, enhancing simulation efficiency. Employing the super-agent concept, we aggregate akin agents within tessellation cells while minimizing perturbations to the dynamics. Our investigation showcases the feasibility of implementing tessellations of varying intricacy through Voronoi diagrams (VDs) while upholding satisfactory quality. Our analysis encompasses diverse city sizes in the United States, spanning small, medium, and large urban areas. The analysis also yields valuable insights: agent count reduction affects results, PCV metric is most sensitive, VD, CBG, and CBGVD exhibit comparable performance with CBGVD excelling in complex features, and city-specific intricacies impact visit duration. 

While applied to varied case studies, our proposed model consistently mirrors the disease spread patterns observed in the FACS-CHARM model. AIC's disease transmission model is intricate, warranting simulations at minute intervals, complemented by hourly synchronization. This simulation encapsulates building occupancy dynamics and myriad activities transpiring within each hour. Furthermore, AIC adeptly manages complex mobility patterns by directly sampling from historical datasets like SafeGraph. Runtime optimization in AIC is anchored in three pivotal strategies: leveraging reduced agent counts through super-agents, tessellating the area of study, and harnessing parallelization techniques.

In the future, our overarching objective is to position the AIC model as a foundational component within a hierarchy of interconnected and cross-validated models. These models will be characterized by their robustness, data-driven nature, and the ability to predict the spread of infectious diseases across diverse spatio-temporal scales. Our ultimate aim is to contribute significantly to effective pandemic prevention strategies. 

The methodology underlying the AIC model, as developed in this manuscript, primarily focuses on representing the finest level of granularity, i.e., individual agents or people. Moving forward, we intend to integrate this fine-grained modeling framework with a coarser modeling approach known as Geo-Graphical Models. These models draw upon insights from fields such as AI, Data Science, Optimization, and Applied Mathematics and Theoretical Engineering \cite{chertkov2021graphical,krechetov2021prediction}. Geo-Graphical Models serve as tools for aggregating individual agents into geo-compartments, which can be thought of as cells within the AIC framework. These compartments could represent various levels of granularity, ranging from households and census tracts to entire cities or states. The resulting aggregated models strike a balance between computational efficiency and accuracy. They will undergo rigorous calibration and validation processes, benchmarked against higher-resolution ABMs (including AIC). Importantly, these models will leverage data-driven approaches, taking full advantage of recent breakthroughs in AI, machine learning, and data science.

Looking ahead, our vision also encompasses collaborative efforts to create more comprehensive datasets. Specifically, we aim to incorporate synthetic data generated by high-resolution models like AIC to train and improve coarser models. This approach not only enhances the modeling pipeline, but also bridges gaps in existing data resources. Our commitment is to continuously advance the field of infectious disease modeling and contribute to more effective pandemic preparedness and response strategies.

\bibliographystyle{unsrtnat}
\bibliography{ms.bib}

\section*{Appendix A: Details of the Agent-In-Cell model}

This appendix serves the purpose of maintaining self-consistency. Here, we provide a comprehensive description of the construction and execution of the Agent-In-Cell (AIC) Agent-Based Model (ABM).

 \subsection{Properties of Geography, Agents and Cells}

\subsubsection{Geography}

 The ABM incorporates two distinct geographical structures:
\begin{enumerate}
\item Tessellation Framework: This entails a tiling strategy applied to a planar geographical map. It involves employing one or more geometric shapes, referred to as tessellation cells, to entirely cover the map without any overlaps or gaps. These tessellation cells can comprise either Census Block Groups (CBGs) or Voronoi Diagram (VD) cells.

\item Point of Interest (POI): This refers to specific geographical coordinates on the map, marked by latitude and longitude. POIs represent precise locations, such as buildings housing institutions like schools, supermarkets, and other such entities.
\end{enumerate}
 
 \subsubsection{Agent Attributes}

Each agent is characterized by the following attributes:
\begin{enumerate}
    
\item Home Tessellation Cell: The tessellation cell in which the agent resides.

\item Home Location: Situated within the agent's home tessellation cell, determined by uniform random sampling based on the cell's population density.

\item Work Destinations List: A list of potential work locations specific to the home tessellation cell. Each work location is assigned a probability, with probabilities summing to unity for a given home tessellation cell. These probabilities are sourced from the known SafeGraph dataset.

\item DayTime CBG: A cell linked to each work location. When the tessellation isn't based on CBGs, the tessellation's overlay with the CBG scheme helps estimate the new distribution of work locations.

\item Assigned Work Location: Randomly assigned based on the list of work locations and associated probabilities.

\item Age: Determined from demographic data of the default tessellation cell (typically a CBG).

\item Occupation: Randomly chosen from a weighted list based on the agent's age, with occupations such as Student, Employee, Medical Staff, Driver, and Unemployed.

\item Group Affiliation: Assigned daily, including the type and size of the group, determined randomly using information from the Open Census Data via SafeGraph. Potential groups comprise Household, Work, Transportation, and Community groups.

\item Tasks: Assigned at the beginning of the day, based on occupation. Random hourly sequences of tasks are time-dependent, including Go to Work, Work, Return Home, Stay Home, Attend Event, Stay in Hospital, and Treat Patients.

\item Agent's Action: Linked to short-duration activities, especially pertinent to the medical model. These actions, assigned before hourly tasks, include options like Sneeze, Contaminate an Object, Physical Contact, Wash Hands, and Touch a Contaminated Object.

\item Medical Status: Agents can be Susceptible, Infected Symptomatic, Infected Asymptomatic, Recovered, or Deceased.

\item Dwell Time: The duration an agent spends at a given Point of Interest (POI). Randomly assigned according to SafeGraph probability data, with time intervals categorized as less than 5 minutes, 5 to 20 minutes, 20 to 60 minutes, 60 to 240 minutes, or greater than 240 minutes.

\item Current Location: Agents can be found at home, work, or a Point of Interest (POI).
\end{enumerate}
 
 \subsubsection{Attributes of a tessellation cell}

 The attributes include:

\begin{enumerate}
    \item List of Destinations: A roster of potential destinations.
\item Visit Frequencies: Frequencies of visits to these destinations.

Current Agents List: An inventory of agents present in the given tessellation cell.

\item Overlapping CBGs: The Census Block Groups (CBGs) with which it shares overlapping boundaries.

\item CBG Overlap Percentage: The proportion of CBGs with which it overlaps.
 \end{enumerate}
 
 \subsubsection{POI Attributes}

The attributes encompass:
\begin{enumerate}
\item Daily Visit Frequencies: Frequencies of visits categorized by the day of the month.
\item Weekly Visit Frequencies: Frequencies of visits categorized by the day of the week.
\item Hourly Visit Frequencies: Frequencies of visits categorized by the hour of the day.
\item POI Building Geometry: The footprint or physical layout of the Point of Interest (POI)'s building.
\item Current Agent List: The present roster of agents associated with the POI.
\item Contamination Level: The current level of contamination attributed to the POI.
 \end{enumerate}
 
 \subsection{Activities}
Agents' daily activities are categorized as either being at their residence, their workplace, or at a Point of Interest (POI).
 \subsubsection{Activities at home and work}
\begin{figure}[h!]
    \centering
    \includegraphics[scale=0.43]{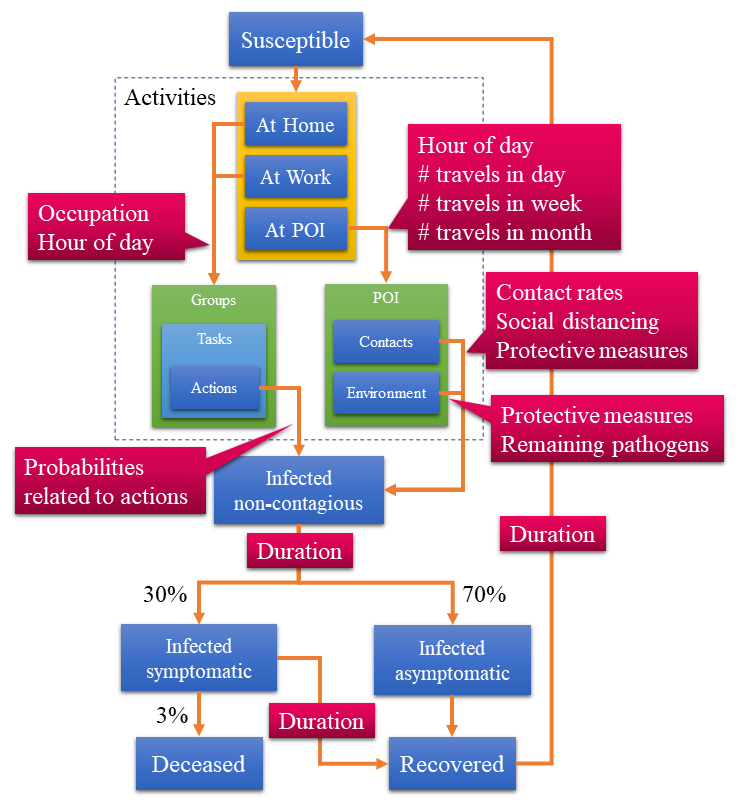}
    \caption{Markov Chain depicting an agent's state space and transitions. Each transition (indicated by an arrow) is characterized by specific rules and a (Poisson) transition rate.}
    \label{fig:abmInfectionStates}
\end{figure}

 \begin{enumerate}
    \item Start of a Day
        \begin{itemize}
            \item Assign all Tasks at the beginning of the day for each agent at random.
            \item Each Task is generated based on the time of the day.
                \begin{itemize}
                    \item Each Task is assigned minimum and maximum starting hours within a day.
                    \item Each Task is assigned minimum and maximum ending hours within a day.
                    \item Each Task is assigned a minimum and maximum probability of occurrence.
                \end{itemize}
            \item Tasks are influenced by the agent's epidemiological status. If the agent is sick, the assigned task becomes "Stay in Hospital".
            \item The generation of tasks also accounts for lockdown and self-quarantine based on externally defined rules.
        \end{itemize}
    \item Hourly Updates
        \begin{itemize}
            \item The agent's actions are generated in accordance with their active task.
            \item The agent's activity groups are updated based on their home and work tessellation cells, dependent on the type and size of the cells.
        \end{itemize}
    \item Daily Epidemiological Status Update
        \begin{itemize}
            \item The agent's epidemiological status is updated at the end of each day, guided by the transition diagram illustrated in Figure \ref{fig:abmInfectionStates}, with associated transition rates.
        \end{itemize}
\end{enumerate}

 \subsubsection{Activities of an Agent at POIs}
 \begin{enumerate}
    \item Decision to Travel to a POI
        \begin{itemize}
            \item The agent's decision to travel to a Point of Interest (POI) depends on the Day of the Month, Day of the Week, and the Hour of the Day.
        \end{itemize}
    \item Agent's Travel from Home/Work to a POI (if not hospitalized)
        \begin{itemize}
            \item Determines the duration of the stay at the POI.
            \item Assesses social distancing by estimating congestion at the POI (using SafeGraph "Number of Visits") and the building's geometry.
            \item Adjusts epidemiological status randomly, adhering to the following rules:
            \begin{itemize}
                \item Infection via Agent-to-Agent Contact
                    \begin{itemize}
                        \item Based on contact rate, social distancing, and mask usage.
                        \item Social distancing and mask usage are translated into a probability of infection for each agent.
                    \end{itemize}
                \item Infection via Agent-to-Environment Contact
                    \begin{itemize}
                        \item Depends on the amount of active pathogens within the environment of the POI.
                        \item Pathogen levels diminish over time, but an infected agent's arrival can lead to re-contamination.
                    \end{itemize}
            \end{itemize}
        \end{itemize}
 \end{enumerate}
 
 \subsection{General Features of the AIC model}

 \begin{itemize}
    \item Global State Update: The global state is updated every minute.
    \item External Agent Arrivals: Agents arriving from outside the geographical domain are randomly assigned infections based on daily information from the John Hopkins' dataset \cite{dong2020interactive}. The number of external daily arrivals is derived from the SafeGraph data.
    \item Infection Trends and Contact Rates: Infection trends and contact rates are reported as outcomes of the Agent-In-Cell (AIC) model.
    \item Data Handling: Due to the size and format of SafeGraph data, which includes patterns and Points of Interest (POIs), this data is read, compressed, and stored on a drive for efficient retrieval during simulation updates.
    \item Geographical Representation: The Federal Information Processing Standards (FIPS) \cite{uscensusbureau2021fips}, as used by the US Census Bureau, are employed for representing geographical regions. All FIPS geographical features of the USA, down to the Census Block Group (CBG) granularity, are pre-processed and stored in a compressed format optimized for swift retrieval.
    \item Geographical Domain Division: The geographical domain is divided into cells, with Voronoi tessellation as the default. Travel time computation relies on OpenStreetMap data and average traffic load.
    \item OpenStreetMap Data Optimization: Given the frequent requests for OpenStreetMap data in its raw form, all potential intersection queries are pre-processed and stored in a compressed, quickly retrievable format.
    \item Epidemiological State Update: An agent's epidemiological state is updated once a day, specifically at day's end, through a parallel processing thread.
\end{itemize}
 
 \subsection{Characteristics of an Agent}
 
 \subsubsection{Occupation}
 
 \begin{itemize}
    \item Service
    Age range: 18-62
    
    \item Student
    Age range: 4-25
    
    \item Doctor
    Age range: 25-70
    
    \item Unemployed
    Age range: 10-81
 \end{itemize}
 
 \subsubsection{Group}
 
 Group sizes are set according to the following rules:
 \begin{itemize}
    \item Average work group size is 10.25
    \item Average family size is 3.9, minimum family size 1, maximum family size 6
    \item Maximum public transportation group size 30
    \item Maximum non-public transportation group size 4
 \end{itemize}
 Transportation group sizes depend on the number of available public transportation seats. Agents are assigned to seats uniformly at random. If there are no seats available, the remaining agents travel without public transportation which can have maximum group size of four. Since the selection of public transportation is stochastic, all public transportation seats won't be occupied. Seat associations with agents is decided according to the following rules
 \begin{itemize}
    \item A set of all transportation seats is defined by $T$
    \item The public and private seats are grouped together. For each agent a private transportation seat is considered because each agent has to reach its destination regardless if public transportation is available or not.
    \item All available transportation seats are added to the set $T$
    \item The set $T$ is shuffled and for each agent a random seat is selected uniformly from the set $T$ 
 \end{itemize}
 
 \subsubsection{Task}
 
Characteristics of each task, determined according to attributes of  agents and the time of the day are:
 \begin{itemize}
    \item Minimum/Maximum starting hour within a day. For instance, going to work can happen from 7am to 9am.
    \item Minimum/Maximum duration of the task in hours
 \end{itemize}
 
 \subsubsection{Action}
 
 The parameters for each action which are not determined by attributes of the agents and the time of the day are
 
 \begin{itemize}
    \item Minimum/Maximum duration: The duration interval of the action.
    \item Minimum/Maximum probability of occurrence: The chance of performing the action by the agent.
    \item Minimum/Maximum effect on other agents: chosen from the $[0,1]$ range depending on how much this action can affect other agents in the same group.
 \end{itemize}
 
 The following parameters are related to the actions of agents while at home or at work. The parameters are independent of agent attributes and the time of the day, however, they are affected by the agent's self-protection measures
 \begin{itemize}
    \item Probability of occurring an action 
    \item Probability for an action to be effective for transmission 
    \item Probability of infection transmission 
 \end{itemize}
 
  \subsubsection{Epidemiological status}
  
  Some transitions in an agent's epidemiological state are time-dependent after the infection starts. The time-dependent transitions are illustrated in Figure \ref{fig:abmInfectionStates}.
\begin{itemize}
    \item Susceptible to Infected non-contagious: This transition occurs at the end of the day when transmission happens.
    \item Infected non-contagious to Infected symptomatic/asymptomatic: This transition occurs after two days, with 70\% of agents becoming asymptomatic.
    \item Infected symptomatic to Dead: Three percent of symptomatic agents succumb to the infection.
    \item Infected symptomatic/asymptomatic to Recovered: This transition occurs between 14 to 16 days.
    \item Recovered to Susceptible: This transition occurs between 55 to 65 days.
\end{itemize}

Additional epidemiological parameters are outlined below:
\begin{itemize}
    \item Self-quarantine: 50\% of agents enter self-quarantine if symptomatic \cite{shamil2021agent}.
    \item Self-protection Level: Ranging from 0 to 1, with a base level of 0.2 that increases based on agents' awareness of symptomatic infections and adoption of protective measures.
    \item Social Events: On average, there are three social events per month per 10,000 people related to the task "Attend event."
    \item Lockdown, Self-quarantine, and Pandemic Awareness: These measures can be initiated at any externally defined time.
\end{itemize}

\end{document}